\documentclass[9pt,journal]{IEEEtran}

\IEEEoverridecommandlockouts        
\usepackage{hyperref}

\title{\LARGE \bf The Observability Radius of Networks}

\author{Gianluca Bianchin, Paolo Frasca, Andrea Gasparri, and
  Fabio~Pasqualetti%
  \thanks{This material is based upon work supported in part by NSF
    awards \#BCS-1430279 and \#ECCS-1405330. {A
      preliminary and abbreviated version of this work will appear in
      the Proceedings of the 2016 American Control Conference
      \cite{GB-PF-AG-FP:16}.} Gianluca Bianchin and Fabio Pasqualetti
    are with the Mechanical Engineering Department, University of
    California at Riverside,
    \href{mailto:gianluca@engr.ucr.edu}{\texttt{gianluca@engr.ucr.edu}},
    \href{mailto:fabiopas@engr.ucr.edu}{\texttt{fabiopas@engr.ucr.edu}}. %
    Paolo Frasca is with the Department of Applied Mathematics,
    University of Twente,
    \href{mailto:p.frasca@utwente.nl}{\texttt{p.frasca@utwente.nl}}. %
    Andrea Gasparri is with the Department of Engineering, Roma Tre
    University,
    \href{mailto:gasparri@dia.uniroma3.it}{\texttt{gasparri@dia.uniroma3.it}}. %
  } }

\usepackage{graphicx,color}
\usepackage{amsmath}
\usepackage{amssymb}
\usepackage{mathrsfs}
\usepackage[vlined,ruled]{algorithm2e}
\usepackage{subfigure}
\usepackage{url}
\usepackage{soul}

\usepackage{booktabs}
\usepackage{array}
\usepackage[table]{xcolor}

\newtheorem{theorem}{Theorem}[section]
\newtheorem{lemma}[theorem]{Lemma}

\newtheorem{corollary}[theorem]{Corollary}

\newtheorem{remark}{Remark}

\newcommand{\map}[3]{#1: #2 \rightarrow #3}
\newcommand{\setdef}[2]{\{#1 \; : \; #2\}}
\newcommand{\subscr}[2]{{#1}_{\textup{#2}}}

\newcommand{\until}[1]{\{1,\dots,#1\}}

\newcommand{\Ker}{\operatorname{Ker}}

\newcommand{\real}{\mathbb{R}}

\newcommand{\transpose}{\mathsf{T}} 
\newcommand{\E}{\operatorname{\mathbb{E}}}

\newcommand{\mc}{\mathcal}

\newcommand{\dd}{{\rm d}}

\DeclareSymbolFont{bbold}{U}{bbold}{m}{n}
\DeclareSymbolFontAlphabet{\mathbbold}{bbold}

\newcommand\oprocendsymbol{\hbox{$\square$}}
\newcommand\oprocend{\relax\ifmmode\else\unskip\hfill\fi\oprocendsymbol}

\renewcommand{\theenumi}{(\roman{enumi})}

\newcommand{\im}{\mathrm{i}}

\renewcommand{\baselinestretch}{0.99}
\begin{document}
\maketitle

\thispagestyle{empty}
\pagestyle{empty}

\begin{abstract}
  This paper studies
  the observability radius of network systems, which measures the
  robustness of a network to perturbations of the edges. We consider
  linear networks, where the dynamics are described by a weighted
  adjacency matrix, and dedicated sensors are positioned at a subset
  of nodes. We allow for perturbations of certain edge weights, with
  the objective of preventing observability of some modes of the
  network dynamics. {To comply with the network setting,
    our} work considers perturbations with a desired sparsity
  structure, thus extending the classic literature on {
    the observability} radius of linear systems. The paper proposes
  two sets of results. First, we propose an optimization framework to
  determine a perturbation with smallest Frobenius norm that renders a
  desired mode unobservable from the existing sensor nodes. Second, we
  study the expected observability radius of { networks
    with given structure and random edge weights}. We provide
  fundamental robustness bounds dependent on the connectivity
  properties of the network and we analytically characterize optimal
  perturbations of line and star networks, showing that line networks
  are inherently more robust than star networks.
\end{abstract}


\section{Introduction}
Networks are broadly used to model engineering, social, and natural
systems. An important property of such systems is their robustness to
contingencies, including failure of components affecting the flow of
information, external disturbances altering individual node dynamics,
and variations in the network topology and weights. 
It remains an
outstanding problem to quantify how different topological features
enable robustness, and to engineer complex networks that remain
operable in the face of arbitrary, and perhaps malicious~perturbations.

Observability of a network guarantees the ability to reconstruct the
state of each node from sparse measurements. While observability is a
binary notion \cite{rek-ych-skn:63}, the degree of observability, akin
to the degree of controllability, can be quantified in different ways,
including the energy associated with the measurements
\cite{FP-SZ-FB:13q,LFC-THS-JL:14}, the novelty of the output signal
\cite{GK-DM-SC:14}, the number of necessary sensor nodes
\cite{AO:14,NM-SZ-CK:14}, and the robustness to removal of
interconnection edges~\cite{MF-AD-UV:14}. A quantitative notion of
observability is preferable over a binary one, as it allows to compare
different observable networks, select optimal sensor nodes, and
identify topological features favoring observability.

In this work we measure robustness of a network based on the size of
the smallest perturbation needed to prevent observability. Our notion
of robustness is motivated by the fact that observability is a generic
property~\cite{wmw:85} and network weights are rarely known without
uncertainty. For these reasons numerical tests to assess observability
may be unreliable and in fact fail to recognize unobservable systems:
instead, our measure of observability robustness can be more reliably
evaluated~\cite{CP:81}. Among our contributions, we highlight
connections between the robustness of a network and its structure, and
we propose an algorithmic procedure to construct optimal
perturbations. Our work finds applicability in network control
problems where the network weights can be changed, in security
applications where an attacker gains control of some network edges,
and in network science for the classification of edges and the design
of robust topologies.


\noindent
\textbf{Related work} Our {study} is inspired by classic
works on the observability radius of dynamical systems
\cite{RE:84,LDB-WSL:86,GH-EJD:04}, defined as the norm of the smallest
perturbation yielding unobservability or, equivalently, the distance
to the nearest unobservable realization. For a linear system described
by { the pair} $(A,C)$, the radius of observability
{ has been classically defined} as
\begin{align*}
  \mu (A, C) = &\min_{\Delta_A, \Delta_C} \left\| 
                 \begin{bmatrix}
                   \Delta_A \\ \Delta_C
                 \end{bmatrix}
               \right\|_2 ,\\[.5em]
  &\text{s.t.} \;\; (A+\Delta_A, C+\Delta_C) \text{ is unobservable}.
\end{align*}
As a known result~\cite{LDB-WSL:86}, the observability radius
satisfies
\begin{align*}
  \mu (A, C) = \min_s \sigma_n \left(
  \begin{bmatrix}
    sI - A \\ C
  \end{bmatrix}
  \right) ,
\end{align*}
where {$\sigma_n$ denotes the smallest singular value},
and $s\in \real$ ($s \in \mathbb{C}$ if complex perturbations are
allowed). The optimal perturbations $\Delta_A$ and $\Delta_C$ are
typically full matrices and, to the best of our knowledge, all
existing results and procedures are not applicable to the case where
the perturbations must satisfy a desired sparsity constraint (e.g.,
see~\cite{MK-DK:09}). This scenario is in fact the relevant one for
network systems, where the nonzero entries of the network matrices $A$
and $C$ correspond to existing network edges, and it would be
undesirable or unrealistic for a perturbation to modify the
interaction of disconnected nodes. An exception is the recent
paper~\cite{MF-AD-UV:14}, where structured perturbations are
considered {in a controllability problem}, yet the
discussion is limited to the removal of edges.




We depart from the literature by requiring the perturbation to be
real, with a desired sparsity pattern, and confined to the network
matrix ($\Delta_C = 0$). Our approach builds on the theory of
\emph{total least squares}~\cite{BDM:94}. With respect to existing
results {on this topic}, our work proposes procedures tailored to
networks, fundamental bounds, and insights into the robustness of
different network topologies.




\noindent
\textbf{Contribution} The contribution of this paper is
threefold. First, we define a metric of network robustness that
captures the resilience of a network system to structural, possibly
malicious, perturbations. Our metric evaluates the distance of a
network from the set of unobservable networks with the same
interconnection structure, and {it extends existing works
  on the observability radius of linear systems.}


Second, we formulate a problem to determine optimal perturbations
(with smallest Frobenius norm) preventing observability. We show that
the problem is not convex, derive optimality conditions, and prove
that any optimal solution solves a nonlinear generalized eigenvalue
problem. Additionally, we propose a numerical procedure based on the
power iteration method to determine (sub)optimal solutions.

Third, we derive a fundamental bound on the expected observability
radius for networks with random weights. 
In particular, we present a class of networks for which the expected
observability radius decays to zero as the network cardinality
increases. Furthermore, we characterize the robustness of line and
star networks. In accordance with recent findings on the role of
symmetries {for the observability and controllability of
  networks}~\cite{AC-MM:14,FP-SZ:14}, we demonstrate that line
networks are inherently more robust than star networks to
perturbations of the edge weights. This analysis shows that our
measure of robustness can in fact be used to compare different network
topologies and guide the design of robust complex systems.

Because the networks we consider are in fact systems with linear
dynamics, our results are generally applicable to linear dynamical
systems. Yet, our setup allows for perturbations with a fixed sparsity
pattern, which may arise from the organization of a network system.


\noindent
\textbf{Paper organization} 
The rest of the paper is organized as follows. 
Section~\ref{sec: setup} contains our network model, the
definition of the network observability radius, and some preliminary
considerations. Section~\ref{sec: frobenius norm} describes our method
to compute network perturbations with smallest Frobenius norm, our
optimization algorithm, and an illustrative example. Our bounds on the
observability radius of random networks are in Section~\ref{sec: line
  star}.
Finally, Section~\ref{sec: conclusion} concludes the~paper.

\section{ The Network Observability Radius}\label{sec: setup}
Consider a directed graph $\mc G := (\mc V, \mc E)$, where
$\mc V := \until{n}$ and $\mc E \subseteq \mc V \times \mc V$ are the
vertex and edge sets, respectively. Let \mbox{$A = [a_{ij}]$} be the
\emph{weighted adjacency matrix} of $\mc G$, where $a_{ij} \in \real$
denotes the weight associated with the edge $(i,j) \in \mc E$
{(representing flow of information from node $j$ to node $i$)}, and $a_{ij} = 0$ whenever
$(i,j) \not\in \mc E$. Let $e_i$ denote the $i$-th canonical vector of
dimension $n$. Let $\mc O = \{ o_1 , \dots, o_p\} \subseteq \mc V$ be
the set of \emph{sensor nodes}, and define the network output matrix
as
                         $
                         C_{\mc O} =
                         \begin{bmatrix}
                           e_{o_1} & \cdots & e_{o_{p}}
                         \end{bmatrix}^\transpose .
                                              $
Let $x_i (t) \in \real$ denote the \emph{state} of node $i$ at time
$t$, and let \mbox{$\map{x}{\mathbb{N}_{\ge 0}}{\real^n}$} be the map
describing the evolution over time of the network state. The network
dynamics are described by the linear discrete-time system
\begin{align}\label{eq: network}
  \begin{split}
    x(t+1) &= A \, x(t), \text{ and }
    y(t) = C_{\mc O} \, x(t),
  \end{split}
\end{align}
where $\map{y}{\mathbb{N}_{\ge 0}}{\real^p}$ is the output of the
sensor nodes $\mc O$.

In this work we characterize structured network perturbations that
prevent observability from the sensor nodes. To this aim, let
\mbox{$\mc H = (\mc V_{\mc H}, \mc E_{\mc H})$} be the \emph{constraint graph}, and
define the set of matrices compatible with $\mc H$ as
\begin{align*}
  \mc A_{\mc H} = \setdef{M}{M \in \real^{|\mc V| \times |\mc V|}, M_{ij} = 0 \text{ if }
  (i,j) \not\in \mc E_{\mc H}}.
\end{align*}
Recall from the eigenvector observability test that the
network~\eqref{eq: network} is observable if and only if there is no
right eigenvector of $A$ that lies in the kernel of $C_{\mc O}$, that
is, $C_{\mc O} x \neq 0$ whenever $x \neq 0$, $A x = \lambda x$, and
$\lambda \in \mathbb{C}$~\cite{TK:80}. In this work we consider and study the
following optimization problem:
\begin{align}\label{eq: minimization problem}
  \begin{array}{lll}
    \min & { \| \Delta \|_\text{F}^2},\\[.5em] 
    \text{s.t.} 
         & (A+\Delta) x =
           \lambda x, & \text{(eigenvalue
                        constraint)},\\[.5em]
         & \| x \|_2 = 1, & \text{(eigenvector
                            constraint)},\\[.5em]
         & C_{\mc O} x = 0, & \text{(unobservability)},\\[.5em]
         & \Delta \in \mc A_{\mc H}, & \text{(structural
                                         constraint)} ,
  \end{array}
\end{align}
where the minimization is carried out over the eigenvector
$x \in \mathbb{C}^n$, the unobservable eigenvalue
$\lambda \in \mathbb{C}$, and the network perturbation
$\Delta \in \real^{n \times n}$. The function
{ $\map{\| \cdot \|_\text{F}}{\real^{n\times n}}{\real_{\ge 0}}$ is the 
Frobenius norm}, 
and $\mc A_{\mc H}$ expresses the desired
sparsity pattern of the perturbation.  It should be observed that (i)~the minimization
problem~\eqref{eq: minimization problem} is not convex because the
variables $\Delta$ and $x$ are multiplied each other in the
eigenvector constraint $(A+\Delta) x = \lambda x$, (ii)~if
$A \in \mc A_{\mc H}$, then the minimization problem is feasible if
and only if there exists a network matrix
$A+ \Delta = \tilde A \in \mc A_{\mc H}$ satisfying the eigenvalue and
eigenvector constraint, and (iii)~if $\mc H = \mc G$, then the
perturbation modifies the weights of the existing edges only. We make
the following assumption:
\begin{itemize}
\item[(A1)] The pair $(A, C_{\mc O})$ is observable. 
\end{itemize}
Assumption (A1) implies that the perturbation $\Delta$ must be nonzero
to satisfy the constraints in~\eqref{eq: minimization problem}. 

{ For the pair $(A, C_{\mc O})$, the \emph{network
    observability radius} is the solution to the optimization
  problem~\eqref{eq: minimization problem}, which quantifies the total
  edge perturbation to achieve unobservability. Different cost
  functions may be of interest and are left as the subject of future
  research.}


The minimization problem~\eqref{eq: minimization problem} can be
solved by two subsequent steps. First, we fix the eigenvalue
$\lambda$, and compute an optimal perturbation that solves the
minimization problem for that $\lambda$. This computation is the topic
of the next section. Second, we search the complex plane for the
optimal $\lambda$ yielding the perturbation with minimum cost. We
observe that (i) the exhaustive search of the optimal $\lambda$ is an
inherent feature of this class of problems, as also highlighted in
prior work~\cite{GH-EJD:04}; (ii) in some cases and for certain
network topologies the optimal $\lambda$ can be found analytically,
{as we do in Section~\ref{sec: line star} for line and
  star networks; and (iii) in certain applications the choice of
  $\lambda$ is guided by the objective of the network
  perturbation, such as inducing unobservability of unstable modes.}

\section{Optimality Conditions and Algorithms for the Network
  Observability Radius}\label{sec: frobenius norm}
In this section we consider problem~\eqref{eq: minimization problem}
with  {\em fixed}
$\lambda$. Specifically, we address the following minimization
problem: given a constraint graph $\mc H$, the network matrix
$A \in \mc A_{\mc G}$, an output matrix $C_{\mc O}$, and a desired
unobservable eigenvalue $\lambda \in \mathbb{C}$, determine a
perturbation $\Delta^* \in \real^{n\times n}$ satisfying
\begin{align}\label{eq: minimization problem F}
  \begin{array}{lll}
    \| \Delta^*\|_\text{F}^2 = \min\limits_{x \in \mathbb{C}^n,  \Delta \in \real^{n\times n}} &  \| \Delta \|_\text{F}^2 ,\\[.7em] 
    \text{s.t.} 
                                                                 & (A+\Delta) x =
                                                                   \lambda
                                                                   x,\\[.5em]
&
                                                                                            \|
                                                                                            x\|_2
    = 1,\\[.5em]
                                                                 &
                                                                   C_{\mc O} x = 0, \\[.5em]
                                                                 &
                                                                   \Delta \in \mc A_{\mc H} .
  \end{array}
\end{align}
From \eqref{eq: minimization problem F}, the value
$\| \Delta^*\|_\text{F}^2$ equals the observability radius of the
network $A$ with sensor nodes $\mc O$, constraint graph $\mc H$, and
fixed unobservable eigenvalue $\lambda$.

\subsection{Optimal network perturbation}
We now shape minimization problem~\eqref{eq: minimization
  problem F} to facilitate its solution. Without affecting generality,
relabel the network nodes such that the sensor nodes set satisfy
\begin{align}\label{eq: labeling}
  \mc O = \until{p} , \text{ so that }
    C_{\mc O} = 
  \begin{bmatrix}
    I_p & 0
  \end{bmatrix}
        .
\end{align}
Accordingly, 
\begin{align}\label{eq: partitioning}
    A =
  \begin{bmatrix}
    A_{11} & A_{12} \\ A_{21} & A_{22}
  \end{bmatrix}
  , \text{ and }
      \Delta =
  \begin{bmatrix}
    \Delta_{11} & \Delta_{12} \\ \Delta_{21} & \Delta_{22}
  \end{bmatrix}
  ,
\end{align}
where $A_{11} \in \real^{p \times p}$,
$A_{12} \in \real^{p \times n-p}$, $A_{21} \in \real^{n-p \times p}$,
and \mbox{$A_{22} \in \real^{n-p \times n-p}$}. Let $V = [v_{ij}]$ be the
unweighted adjacency matrix of $\mc H$, where $v_{ij} = 1$ if
$(i,j) \in \mc E_{\mc H}$, and $v_{ij} = 0$ otherwise. Following the
partitioning of $A$ in~\eqref{eq: partitioning}, let
\begin{align*}
  V =
  \begin{bmatrix}
    V_{11} & V_{12}\\
    V_{21} & V_{22}
  \end{bmatrix}
             .
\end{align*}

We perform the following three simplifying steps.

\noindent
\emph{(1--Rewriting the structural constraints)} Let $B = A + \Delta$,
and notice that
$\| \Delta\|_\text{F}^2 = \sum_{i = 1}^n \sum_{j = 1}^n \; (b_{ij} -
a_{ij})^2$.
Then, the minimization problem~\eqref{eq: minimization problem F} can
equivalently be rewritten restating the constraint
$\Delta \in \mc A_{\mc H}$, as in the following:
\begin{align*}
  \| \Delta\|_\text{F}^2 = \| B - A\|_\text{F}^2 = \sum_{i = 1}^n
  \sum_{j = 1}^n \;  (b_{ij} - a_{ij})^2
  v_{ij}^{-1} .
\end{align*}
Notice that $\| \Delta\|_\text{F}^2 = \infty$ whenever $\Delta$ does
not satisfy the structural constraint, that is, when $v_{ij} = 0$ and
$b_{ij} \neq a_{ij}$.


\noindent
\emph{(2--Minimization with real variables)} Let
$\lambda = \lambda_{\Re} + \im \lambda_{\Im}$, where $\im$ denotes the
imaginary unit. Let
\begin{align*}
  x_\Re =
  \begin{bmatrix}
    x_\Re^1 \\ x_\Re^2
  \end{bmatrix} , \text{ and }
  x_\Im =
  \begin{bmatrix}
    x_\Im^1 \\ x_\Im^2
  \end{bmatrix}
  ,
\end{align*}
denote the real and imaginary parts of the eigenvector $x$, with
\mbox{$x_\Re^1\in \real^{p}$}, $x_\Im^1 \in \real^{p}$,
$x_\Re^2 \in \real^{n-p}$, and $x_\Im^2 \in \real^{n-p}$.
\begin{lemma}{\bf \emph{(Minimization with real eigenvector constraint)}}\label{lemma:
    eq decoupling}
  The constraint $(A + \Delta) x = \lambda x$ can equivalently be
  written as
\begin{align}\label{eq: eig real imag}
  \begin{split}
    (A+\Delta - \lambda_{\Re} I) x_\Re &= - \lambda_{\Im} x_\Im,\\
    (A+\Delta - \lambda_{\Re} I) x_\Im &= \lambda_{\Im} x_\Re  .
  \end{split}
\end{align}
\end{lemma}
\begin{IEEEproof}
  By considering separately the real and imaginary part of the
  eigenvalue constraint, we have
  $(A + \Delta) x = \lambda_{\Re} x + \im \lambda_{\Im} x$ and
  $(A + \Delta) \bar{x} = \lambda_{\Re} \bar{x} - \im \lambda_{\Im}
  \bar{x}$,
  where $\bar x$ denotes the complex conjugate of $x$.  Notice that
  \begin{align*}
    \underbrace{(A + \Delta)  (x + \bar{x})}_{(A+ \Delta) 2 
    x_{\Re}}  = \underbrace{(\lambda_{\Re} + \im \lambda_{\Im}) x
    + (\lambda_{\Re} - \im \lambda_{\Im}) \bar x}_{2 \lambda_{\Re}
    x_{\Re} - 2 \lambda_{\Im} x_{\Im}} ,
  \end{align*}
  and, analogously,
  \begin{align*}
    \underbrace{(A + \Delta)  (x - \bar{x})}_{(A+ \Delta) 2 \im
    x_{\Im}}  = \underbrace{(\lambda_{\Re} + \im \lambda_{\Im}) x
    - (\lambda_{\Re} - \im \lambda_{\Im}) \bar x}_{2 \im \lambda_{\Re}
    x_{\Im} + 2 \im \lambda_{\Im} x_{\Re}} ,
  \end{align*}
    which concludes the proof.
\end{IEEEproof}
Thus, the problem~\eqref{eq: minimization problem F} can be solved
over real variables only.

\noindent
\emph{(3--Reduction of dimensionality)} The constraint $C_{\mc O} x = 0$
and equation~\eqref{eq: labeling} imply that $x_\Re^1 = x_\Im^1 = 0$.
Thus, in the minimization problem~\eqref{eq: partitioning} we set
$\Delta_{11} = 0$, $\Delta_{21} = 0$, and consider the minimization
variables $x_\Re^2$, $x_\Im^2$, $\Delta_{12}$, and $\Delta_{22}$.

These simplifications lead to the following result.
\begin{lemma}{\bf \emph{(Equivalent minimization
      problem)}}\label{lemma: eq prob}
    Let
  \begin{align}\label{eq: bar system def}
    \begin{split}
    \bar{A} &= 
    \begin{bmatrix}
      A_{12} \\ A_{22}
    \end{bmatrix}
    ,
    \bar{\Delta} = 
    \begin{bmatrix}
      \Delta_{12} \\
      \Delta_{22}
    \end{bmatrix}
    ,
    \bar{M} = 
    \begin{bmatrix}
      0_{p\times n-p} \\
      \lambda_\Im I_{n-p}
    \end{bmatrix}
    , \\
    \bar{N} &= 
    \begin{bmatrix}
      0_{p\times n-p} \\
      \lambda_\Re I_{n-p}
    \end{bmatrix},
    \bar V =
    \begin{bmatrix}
      V_{12} \\ V_{22}
    \end{bmatrix},
    \text{ and }
    \bar B = \bar A + \bar \Delta.
    \end{split}
  \end{align}
  The following minimization problem is equivalent to~\eqref{eq:
    minimization problem F}:
    \vspace{-2mm}
  \begin{align}\label{eq: reduced problem}
    \begin{array}{lll}
      \| \bar{\Delta}^*\|_\text{F}^2 \; = \min\limits_{\bar B,
      x_{\Re}^2, x_{\Im}^2 }  & \displaystyle\sum_{i = 1}^n \; {\sum_{j = 1}^{n-p}} \;
                                (\bar b_{ij} - \bar a_{ij  } )^2 v_{ij}^{-1} ,\\[1.2em] 
      \text{s.t.} 
                                      & 
                                        \begin{bmatrix}
                                          \bar B-\bar N & \bar M\\
                                          -\bar M & \bar B-\bar N
                                        \end{bmatrix}
                                                   \begin{bmatrix}
                                                     x_{\Re}^2 \\ x_{\Im}^2
                                                   \end{bmatrix}
      =
      0 ,\\[1.0em]
      &
      \left\|
      \begin{bmatrix}
        x_{\Re}^2 \\ x_{\Im}^2
      \end{bmatrix}
      \right\|_2 = 1.
      \end{array}
  \end{align}
\end{lemma}
\smallskip

The minimization problem~\eqref{eq: reduced problem} belongs to the
class of \emph{(structured) total least squares} problems, which arise
in several estimation and identification problems in control theory
and signal processing.
%
Our approach is inspired by~\cite{BDM:94}, with the difference that we
focus on real perturbations $\Delta$ and complex eigenvalue $\lambda$:
this constraint leads to different optimality conditions and
algorithms. {Let $A \otimes B$ denote the Kronecker
  product between the matrices $A$ and $B$, and
  $\text{diag}(d_1,\dots,d_n)$ the diagonal matrix with scalar entries
  $d_1,\dots, d_n$.} We now derive the optimality conditions for the
problem~\eqref{eq: reduced problem}.

\begin{theorem}{\bf \emph{(Optimality conditions)}}\label{thm: nec
    conditions}
  Let $x_\Re^*$, and $x_\Im^*$ be a solution to the
  minimization problem~\eqref{eq: reduced problem}. Then,
  \begin{align}\label{eq: optimality}
    \begin{split}
    \underbrace{
    \begin{bmatrix}
      \bar A - \bar N & \bar M\\
     - \bar M & \bar A - \bar N
    \end{bmatrix}}_{\tilde A}
                \underbrace{
                \begin{bmatrix}
                  x_\Re^* \\
                  x_\Im^*
                \end{bmatrix}}_{x^*}
    &= \sigma
      \underbrace{
      \begin{bmatrix}
        S_x & T_x\\
        T_x & Q_x
      \end{bmatrix}}_{D_x}
              \underbrace{
              \begin{bmatrix}
                y_1\\
                y_2
              \end{bmatrix}}_{y^*}
              ,\\
    \underbrace{
    \begin{bmatrix}
      \bar A - \bar N & \bar M\\
     - \bar M & \bar A - \bar N
    \end{bmatrix}^\transpose}_{\tilde A^\transpose}
                            \underbrace{
                \begin{bmatrix}
                y_1\\
                y_2
                \end{bmatrix}}_{y^*}
    &= \sigma
      \underbrace{
      \begin{bmatrix}
        S_y & T_y\\
        T_y & Q_y
      \end{bmatrix}}_{D_y}
              \underbrace{
              \begin{bmatrix}
                  x_\Re^* \\
                  x_\Im^*
              \end{bmatrix}}_{x^*}
    ,
    \end{split}
  \end{align}
  for some $\sigma > 0$ and $y^* \in \real^{2n}$ with $\| y^*\| = 1$,
  and where
  \begin{align}\label{eq: def S,T,Q}
    \begin{split}
      D_1 &=
      \text{diag}(v_{11},\dots,v_{1n},v_{21},\dots,v_{2n},\dots,v_{n1},\dots,v_{nn}
      ),\\[.3em]
      D_2 &= \text{diag}(v_{11},\dots,v_{n1},v_{12},\dots,v_{n2}
      ,\dots,v_{1n} ,\dots,v_{nn}
      ), \\[.3em]
      S_x &= (I \otimes x_\Re^*)^\transpose D_1 (I \otimes x_\Re^*) ,
      \;T_x = (I \otimes x_\Re^*)^\transpose D_1 (I \otimes x_\Im^*) ,\\[.3em]
      Q_x &= (I \otimes x_\Im^*)^\transpose D_1 (I \otimes x_\Im^*) ,
      \;S_y = (I \otimes y_1)^\transpose D_2 (I \otimes y_1) ,\\[.3em]
      T_y &= (I \otimes y_1)^\transpose D_2 (I \otimes y_2) ,
      \;Q_y = (I \otimes y_2)^\transpose D_2 (I \otimes y_2) .
    \end{split}
  \end{align}
\end{theorem}
\smallskip
\begin{IEEEproof}
  We adopt the method of Lagrange multipliers to derive optimality
  conditions for the problem~\eqref{eq: reduced problem}. The
  Lagrangian is
  \begin{align}\label{eq: Lagrangian}
  \nonumber
    \mc L &(\bar B, x_\Re^2, x_\Im^2, \ell_1, \ell_2 , \rho) = \sum_{i}
    \sum_{j} (\bar b_{ij} - \bar a_{ij})^2 v_{ij}^{-1} \\ 
      \nonumber
      &+ \ell_1^\transpose
    ((\bar B-\bar N) x_\Re^2 + \bar M x_\Im^2) + \ell_2^\transpose
    ((\bar B-\bar N) x_\Im^2 -\bar M x_\Re^2) \\ &+ \rho (1 - x_\Re^{2\transpose}
    x_\Re^{2} - x_\Im^{2\transpose}x_\Im^{2}),
  \end{align}
  where $\ell_1\in \real^{n}$, $\ell_2\in \real^{n}$, and
  $\rho\in\real$ are Lagrange multipliers.  By equating the partial
  derivatives of $\mc L$ to zero we obtain
  \begin{align}
    \frac{\partial \mc L}{\partial b_{ij}} = 0 &\Rightarrow -2 (\bar
                                                 a_{ij} - \bar b_{ij}) v_{ij}^{-1} + \ell_{1i} x_{\Re j}^2 + \ell_{2i} x_{\Im
                                                 j}^2 = 0, \label{eq:
                                                 partial b}\\
    \frac{\partial \mc L}{\partial x_{\Re}^2} = 0 &\Rightarrow
                                                    \ell_1^\transpose
                                                    (\bar B-\bar N) -
                                                    \ell_2^\transpose
                                                    \bar M -2 \rho
                                                    x_\Re^{2\transpose}
                                                    = 0
                                                    , \label{eq:
                                                    partial xr}\\
    \frac{\partial \mc L}{\partial x_{\Im}^2} = 0 &\Rightarrow   \ell_1^\transpose
                                                    \bar M +
                                                    \ell_2^\transpose
                                                    (\bar B-\bar N) -2 \rho
                                                    x_\Im^{2\transpose}
                                                    = 0 , \label{eq:
                                                    partial XI}\\
    \frac{\partial \mc L}{\partial \ell_1} = 0 &\Rightarrow   (\bar B-\bar N)
                                                 x_\Re^2 + \bar M
                                                 x_\Im^2 = 0
                                                 , \label{eq: partial l1}\\
    \frac{\partial \mc L}{\partial \ell_2} = 0 &\Rightarrow  (\bar B-\bar N)
                                                 x_\Im^2 -\bar M
                                                 x_\Re^2 = 0  , \label{eq: partial l2}\\
    \frac{\partial \mc L}{\partial \rho} = 0 &\Rightarrow    x_\Re^{2\transpose}
    x_\Re^{2} + x_\Im^{2\transpose}x_\Im^{2} = 1.
  \end{align}

  Let $L_1 = \text{diag}(\ell_1)$, $L_2 = \text{diag}(\ell_2)$,
  $X_\Re = \text{diag}(x_\Re^2)$, $X_\Im = \text{diag}(x_\Im^2)$.
  After including the factor $2$ into the multipliers, equation
 ~\eqref{eq: partial b} can be written in matrix form as
  \begin{align}\label{eq: partial B}
    \bar A - \bar B = L_1 \bar V X_\Re + L_2 \bar V X_\Im .
  \end{align}
  Analogously, equations~\eqref{eq: partial xr} and~\eqref{eq: partial
    XI} can be written as
  \begin{align}\label{eq: partial xr+xi}
    \begin{bmatrix}
      \ell_1^\transpose & \ell_2^\transpose
    \end{bmatrix}
                          \begin{bmatrix}
                            \bar B-\bar N & \bar M\\
                            -\bar M & \bar B-\bar N
                          \end{bmatrix}
                                      -2 \rho
                                      \begin{bmatrix}
                                        x_\Re^{2\transpose} & x_\Im^{2\transpose}
                                      \end{bmatrix}
                                                              = 0  ,
  \end{align}
  From equation~\eqref{eq: partial xr+xi} we have
  \begin{align*}
    \begin{bmatrix}
      \ell_1^\transpose & \ell_2^\transpose
    \end{bmatrix}
                          \underbrace{
                          \begin{bmatrix}
                            \bar B-\bar N & \bar M\\
                            -\bar M & \bar B-\bar N
                          \end{bmatrix}
                                      \begin{bmatrix}
                                        x_\Re^{2} \\ x_\Im^{2}
                                      \end{bmatrix}}_{=0 \text{ due to
   ~\eqref{eq: partial l1} and~\eqref{eq: partial l2}}}
                                      -2 \rho
                                      = 0 ,
  \end{align*}
  from which we conclude $\rho = 0$. By combining~\eqref{eq: partial
    l1} and~\eqref{eq: partial B} (respectively,~\eqref{eq: partial l2}
  and~\eqref{eq: partial B}) we obtain
  \begin{align*}
    (\bar A - \bar N) x_\Re^2 + \bar M x_\Im^2 &= \left( L_1 \bar V X_\Re + L_2 \bar V X_\Im
                                                 \right) x_\Re^2 , \\
    (\bar A - \bar N) x_\Im^2 - \bar M x_\Re^2 &= \left( L_1 \bar V X_\Re + L_2 \bar V X_\Im
                                                 \right) x_\Im^2 .
  \end{align*}
  Analogously, by combining~\eqref{eq: partial xr} and~\eqref{eq:
    partial B}, \eqref{eq: partial XI} and~\eqref{eq: partial B}, we
  obtain
  \begin{align*}
    \ell_1^\transpose (\bar A - \bar N) - \ell_2^\transpose \bar M &=
    \ell_1^\transpose \left( L_1 \bar V X_\Re + L_2 \bar V X_\Im
    \right) , \\
    \ell_2^\transpose (\bar A - \bar N) + \ell_1^\transpose \bar M &=
    \ell_2^\transpose \left( L_1 \bar V X_\Re + L_2 \bar V X_\Im
    \right) .
  \end{align*}
%
  Let
  $\sigma = \sqrt{\ell_1^\transpose \ell_1 + \ell_2^\transpose
    \ell_2}$
  and observe that $\sigma$ cannot be zero.  Indeed, due to Assumption
  (A1), the optimal perturbation can not be zero; thus, the first
  constraint in~\eqref{eq: reduced problem} must be active and the
  corresponding multiplier must be nonzero.  Then, we can define
  $y_1 = \ell_1/\sigma$ and $y_2 = \ell_2/\sigma$ and we can verify
  that
  \begin{align*}
    \left( L_1 \bar V X_\Re + L_2 \bar V X_\Im
    \right) x_\Re^2 &= \sigma \left( S_x y_1 + T_x y_2 \right), \\
    \left( L_1 \bar V X_\Re + L_2 \bar V X_\Im
    \right) x_\Im^2 &= \sigma \left( T_x y_1 + Q_x y_2 \right), 
  \end{align*}
  and
  \begin{align*}
    \sigma \left( y_1^\transpose (\bar A - \bar N) - y_2^\transpose \bar M
    \right) &= 
              \ell_1^\transpose \left( L_1 \bar V X_\Re + L_2 \bar V X_\Im
              \right) \\ & = \sigma^2 \left( S_y x_\Re^2 + T_y x_\Im^2  \right)^\transpose, 
\\
    \sigma \left( y_2^\transpose (\bar A - \bar N) + y_1^\transpose \bar M
    \right) &= \ell_2^\transpose \left( L_1 \bar V X_\Re + L_2 \bar V X_\Im
              \right) \\ &= \sigma^2 \left( T_y x_\Re^2 + Q_y x_\Im^2  \right)^\transpose,
  \end{align*}
  which conclude the proof.
\end{IEEEproof}
Note that equations~\eqref{eq: optimality} may admit multiple
solutions, and that every solution to~\eqref{eq: optimality} yields a
network perturbation that satisfies the constraints in the
minimization problem~\eqref{eq: reduced problem}. We now present the
following result to compute perturbations.

\begin{corollary}{\bf \emph{(Minimum norm perturbation)}}\label{corollary:
    perturbation}
  Let $\Delta^*$ be a solution to~\eqref{eq: minimization problem
    F}. Then, $\Delta^* = [0^{n \times p} \; \bar \Delta^*]$, where
  \begin{align*}
    \bar \Delta^* = - \sigma \left( \text{diag}(y_1) \bar V
    \text{diag}(x_\Re^*) - \text{diag}(y_2) \bar V \text{diag}(x_\Im^*)
    \right),
  \end{align*}
  and $x^*_\Re$, $x^*_\Im$, $y_1$, $y_2$, $\sigma$ satisfy the
  equations~\eqref{eq: optimality}. Moreover,
  \begin{align*}
    \| \Delta \|_\text{F}^2 &= \sigma^2 x^{*\transpose} D_y x^* =
                              \sigma x^{*\transpose} \tilde A^\transpose y^* \le \sigma \|
                              \tilde A \|_\text{F} .
  \end{align*}
\end{corollary}
\smallskip
\begin{IEEEproof}
  The expression for the perturbation $\Delta^*$ comes from Lemma
  ~\ref{lemma: eq prob} and~\eqref{eq: partial B}, and the fact that
  $L_1 = \sigma \, \text{diag}(y_1)$,
  \mbox{$L_2 = \sigma \, \text{diag}(y_2)$}. To show the second part notice
  that
  \begin{align*}
    \| \Delta \|_\text{F}^2 &= \| A - B\|_\text{F}^2 = \| L_1 \bar V X_\Re + L_2 \bar V X_\Im \|_\text{F}^2 \\
                            &= \sigma^2 \sum_i \sum_j
                              \left( y_{1i}^2
                              x_{\Re j}^2 +
                              y_{2i}^2 x_{\Im
                              j}^2\right)
                              v_{ij} \\
                            &= \sigma^2
                              x^{*\transpose}
                              D_y x^* =
                              \sigma x^{*\transpose} \tilde
                              A^\transpose y^* ,
  \end{align*}
  where the last equalities follow from~\eqref{eq:
    optimality}. Finally, the inequality follows from
  $\| x^* \|_2 = \|x^*\|_\text{F}=\| y^* \|_2 = \|y^*\|_\text{F} =
  1$.
\end{IEEEproof}

To compute a triple $(\sigma, x^*, y^*)$ satisfying the condition in
Theorem~\ref{thm: nec conditions}, observe that \eqref{eq: optimality}
can be written in matrix form as
\begin{align}\label{eq: generalized eig}
  \underbrace{
  \begin{bmatrix}
      0 & \tilde A^\transpose \\
    \tilde A & 0
  \end{bmatrix}}_{H}
        \underbrace{
        \begin{bmatrix}
          x \\ y
        \end{bmatrix}}_{z}
  =
  \bar \sigma
  \underbrace{
  \begin{bmatrix}
      D_y & 0 \\
    0 & D_x
  \end{bmatrix}}_{D}
          \underbrace{
          \begin{bmatrix}
            x \\ y
          \end{bmatrix}}_{z}
  .
\end{align}
\begin{lemma}{\bf \emph{(Equivalence between Theorem~\ref{thm: nec
        conditions} and~\eqref{eq: generalized eig})}}\label{lemma:
    equivalence}
  Let $(\sigma, x , y)$, with $x \neq 0$, solve~\eqref{eq: generalized
    eig}. Then, $\sigma \neq 0$ and $y \neq 0$, and the triple
  $((\alpha \beta )^{-1}\sigma, \alpha x, \beta y)$, with
  $\alpha = \text{sgn} (\sigma)\| x \|^{-1}$ and
  $\beta = \| y \|^{-1}$, satisfies the conditions in Theorem~\ref{thm:
    nec conditions}.
\end{lemma}
\begin{IEEEproof}
  Because $x \neq 0$ and $\tilde A$ has full column rank due to
  Assumption (A1), it follows $\sigma \neq 0$ and $y \neq 0$. Let
  $D_x$ and $D_y$ be as in~\eqref{eq: optimality}. Notice that
  $D_{\alpha x} = \alpha^2 D_{ x}$ and
  $D_{\beta y} = \beta^2 D_{ y}$. Notice that
  $(\alpha \beta )^{-1} \sigma > 0$. We have
  \begin{align*}
    \tilde A \alpha x &= \frac{ \sigma}{\alpha \beta} \alpha^2 D_x
    \beta y = \alpha \sigma D_x y,\\
    \tilde A^{\transpose} \beta y &=  \frac{ \sigma}{\alpha \beta}
                                    \beta^2 D_y \alpha x = \beta 
                                    \sigma D_y x ,
  \end{align*}
  which concludes the proof.
\end{IEEEproof}
Lemma~\ref{lemma: equivalence} shows that a (sub)optimal network
perturbation can in fact be constructed by solving equations
\eqref{eq: generalized eig}. It should be observed that, if the
matrices $S_x$, $T_x$, $Q_x$, $S_y$, $T_y$, and $Q_y$ were constant,
then~\eqref{eq: generalized eig} would describe a generalized
eigenvalue problem, thus a solution $(\bar \sigma, z)$ would be a pair
of generalized eigenvalue and eigenvector. These facts will be
exploited in the next section to develop a heuristic algorithm to
compute a (sub)optimal network perturbation.

\begin{remark}{\bf \emph{(Smallest network perturbation with respect to
      the unobservable eigenvalue)}}\label{remark: smallest perturbation}
  In the minimization problem~\eqref{eq: minimization problem F} the
  size of the perturbation $\Delta^*$ depends on the desired
  eigenvalue $\lambda$, and it may be of interest to characterize the
  unobservable eigenvalue
 \mbox{$\lambda^* = \lambda_\Re^* + \im \lambda_\Im^*$} yielding the
  smallest network perturbation that prevents observability. To this
  aim, we
equate to zero the derivatives of the Lagrangian~\eqref{eq:
    Lagrangian} with respect to $\lambda_\Re$ and $\lambda_\Im$ to
  obtain
  \begin{align*}
    \frac{\partial \mc L}{\partial \lambda_\Re} &= 0 \Rightarrow
                                                  \ell_1^\transpose
                                                  \begin{bmatrix}
                                                    0_p \\ x_\Re^2
                                                  \end{bmatrix}
    +
    \ell_2^\transpose
    \begin{bmatrix}
      0_p \\ x_\Im^2
    \end{bmatrix}
    = 0,\\[.3em]
    \frac{\partial \mc L}{\partial \lambda_\Im} &= 0 \Rightarrow
                                                  \ell_1^\transpose
                                                  \begin{bmatrix}
                                                    0_p \\ x_\Im^2
                                                  \end{bmatrix}
    -
    \ell_2^\transpose
    \begin{bmatrix}
      0_p \\ x_\Re^2
    \end{bmatrix}
    = 0.
  \end{align*}
  The above conditions clarify that, for the perturbation $\Delta$ to
  be of the smallest size with respect to $\lambda$, the Lagrange
  multipliers $\ell_1$ and $\ell_2$, and the vectors $x_\Re^2$ and
  $x_\Im^2$ must verify an orthogonality condition. \oprocend
\end{remark}

\begin{remark}{\bf \emph{(Real unobservable
      eigenvalue)}}\label{remark: real eigenvalues}
  When the unobservable eigenvalue $\lambda$ in~\eqref{eq:
    minimization problem F} is real, the optimality conditions in
  Theorem~\ref{thm: nec conditions} can be simplified to
  \begin{align*}
    (\bar{A} - \bar{N}) x_{\Re} = \sigma S_x y_1, \text{ and } 
    (\bar{A} - \bar{N}) y_1 = \sigma S_y x_{\Re} .
  \end{align*}
  The generalized eigenvalue equation~\eqref{eq: generalized eig}
  becomes
  \begin{align*}
    \begin{bmatrix}
      0 & \bar{A}^\transpose - \bar{N}^\transpose\\
      \bar A - \bar N & 0
    \end{bmatrix}
                        \begin{bmatrix}
                          y_1 \\ x_\Re
                        \end{bmatrix}
    =
    \sigma
    \begin{bmatrix}
      S_x & 0\\
      0 & S_y
    \end{bmatrix}
          \begin{bmatrix}
            y_1 \\ x_\Re
          \end{bmatrix} ,
  \end{align*}
  and the optimality conditions with respect to the unobservable
  eigenvalue $\lambda$ (see Remark~\ref{remark: smallest
    perturbation}) simplify to $\ell_1^\transpose
                                                  \begin{bmatrix}
                                                    0_p \\ x_\Re^2
                                                  \end{bmatrix} = 0$.
\oprocend
 \end{remark}

\subsection{A heuristic procedure to compute structural perturbations}
In this section we propose an algorithm to find a solution to the set
of nonlinear equations~\eqref{eq: generalized eig}, and thus to find a
(sub)optimal solution to the minimization problem~\eqref{eq:
  minimization problem F}. Our procedure is motivated by~\eqref{eq:
  generalized eig} and Corollary~\ref{corollary: perturbation}, and it
consists of fixing a vector $z$, computing the matrix $D$, and
approximating an eigenvector associated with the smallest generalized
eigenvalue of the pair $(H,D)$. Because the size of the perturbation
is bounded by the generalized eigenvalue $\sigma$ as in
Corollary~\ref{corollary: perturbation}, we adopt an iterative
procedure based on the \emph{inverse iteration} method for the
computation of the smallest eigenvalue of a matrix
\cite{LNT-DB:97}. We remark that our procedure is heuristic, because
\eqref{eq: generalized eig} is in fact a nonlinear generalized
eigenvalue problem due to the dependency of the matrix $D$ on the
eigenvector $z$. To the best of our knowledge, no complete algorithm
is known for the solution of~\eqref{eq: generalized eig}. We start by
characterizing certain properties of the matrices $H$ and $D$, which
will be used to derive our algorithm. Let
\begin{align*}
  \text{spec} (H,D) = \setdef{\lambda \in \mathbb{C}}{ \text{det} (H -
  \lambda D) =0 },
\end{align*}
and recall that the pencil $(H,D)$ is regular if the determinant
\mbox{$\text{det} (H - \lambda D)$} does not vanish for some value of
$\lambda$, { see}~\cite{GHG-CFVL:12}. Notice that, if $(H,D)$ is not regular,
then $\text{spec} (H,D) = \mathbb{C}$.

\begin{lemma}{\bf \emph{(Generalized eigenvalues of
      $(H,D)$)}}\label{lemma: generalized eig}
  Given a vector {$z \in \real^{4n - 2p}$}, define the
  matrices $H$ and $D$ as in~\eqref{eq: generalized eig}. Then,
  \begin{enumerate}
  \item $0 \in \text{spec}(H,D)$;
  \item if $\lambda \in \text{spec}(H,D)$, then $-\lambda \in
    \text{spec}(H,D)$; and
  \item if $(H,D)$ is regular, then $\text{spec}(H,D) \subset \real$.
  \end{enumerate}
\end{lemma}
\smallskip
\begin{IEEEproof}
  Statement (i) is equivalent to $\tilde A x = 0$ and
  $\tilde A^\transpose y = 0$, for some vectors $x$ and $y$. Because
  $\tilde A^\transpose \in \real^{(2n-2p) \times 2n}$ with $p \ge 1$,
  the matrix $\tilde A^\transpose$ features a nontrivial null
  space. Thus, the two equations are satisfied with $x = 0$ and
  $y \in \Ker (\tilde A^\transpose)$, and the statement follows.

  To prove statement (ii) notice that, due to the block structure of
  $H$ and $D$, if the triple $(\lambda, \bar x, \bar y)$ satisfies the
  generalized eigenvalue equations
  $\tilde A^\transpose \bar y = \lambda D_y \bar x$ and
  $\tilde A \bar x = \lambda D_x \bar y$, so does
  $(-\lambda, \bar x, - \bar y)$.


  To show statement (iii), let $\text{Rank}(D) = k \le n$, and notice
  that the regularity of the pencil $(H,D)$ implies $H \bar z \neq 0$
  whenever $D \bar z = 0$ and $\bar z \neq 0$. Notice that $(H,D)$ has
  $n-k$ infinite eigenvalues~\cite{GHG-CFVL:12} because
  $H \bar z = \lambda D \bar z = \lambda \cdot 0$ for every nontrivial
  $\bar z \in \Ker(D)$.
  Because $D$ is symmetric, it admits an orthonormal basis of
  eigenvectors. Let $V_1 \in \real^{n \times k}$ contain the
  orthonormal eigenvectors of $D$ associated with its nonzero
  eigenvalues, let $\Lambda_D$ be the corresponding diagonal matrix of
  the eigenvalues, and let $T_1 = V_1 \Lambda_D^{-1/2}$.  Then,
  $T_1^\transpose D T_1 = I$. Let $\tilde H = T_1^\transpose H T_1$,
  and notice that $\tilde H$ is symmetric. Let
  $T_2 \in \real^{k \times k}$ be an orthonormal matrix of the
  eigenvectors of $\tilde H$. Let $T = T_1T_2$ and note that
    $
    T^\transpose H T = \Lambda \text{ and } T^\transpose D T = I ,
    $
  where $\Lambda$ is a diagonal matrix. To conclude, consider the
  generalized eigenvalue problem $H \bar z = \lambda D \bar z$. Let
  $\bar z = T \tilde z$. Because $T$ has full column rank $k$, we have
    $
    T^\transpose H T \tilde z = \Lambda \tilde z = \lambda
    T^\transpose D T \tilde z = \lambda \tilde z,
    $
  from which we conclude that $(H,D)$ has $k$ real eigenvalues.
\end{IEEEproof}

Lemma~\ref{lemma: generalized eig} implies that the inverse iteration
method
is not directly applicable to~\eqref{eq: generalized eig}. In fact,
the zero eigenvalue of $(H,D)$ leads the inverse iteration to
instability, while the presence of eigenvalues of $(H,D)$ with equal
magnitude may induce non-decaying oscillations in the solution
vector. To overcome these issues, we employ a shifting mechanism as
detailed in Algorithm~\ref{alg: IPI}, where the eigenvector $z$ is
iteratively updated by solving the equation $(H-\mu D) z_{k+1}= D z_k$
until a convergence criteria is met.  Notice that (i) {the
  eigenvalues of $(H- \mu D, D)$ are shifted with respect to the
  eigenvalues of $(H,D)$, that is, if $\sigma \in \text{spec}(H, D)$,
  then
  $\sigma - \mu \in \text{spec}(H - \mu D,
  D)$,\footnote{{
      To see this, let $\sigma$ be an eigenvalue of $(H, D)$, that is,
      $Hx = \sigma D x$. Then,
      $(H - \mu D) x = H x - \mu D x = \sigma D x - \mu D x = (\sigma
      - \mu) D x$.
      That is $(H - \mu D) x = (\sigma - \mu) D x$ thus $\sigma - \mu$
      is an eigenvalue of $(H - \mu D, D)$.  }}} (ii)~the pairs
$(H- \mu D, D)$ and $(H,D)$ share the same eigenvectors, and (iii)~by
selecting
$\mu = \psi \cdot \min \setdef{\sigma \in \text{spec}(H,D)}{\sigma >
  0}$,
the pair $(H- \mu D, D)$ has nonzero eigenvalues with distinct
magnitude. Thus, Algorithm~\ref{alg: IPI} estimates the eigenvector
$z$ associated with the smallest nonzero eigenvalue $\sigma$ of
$(H,D)$, and converges when $z$ and $\sigma$ also satisfy
equations~\eqref{eq: generalized eig}. The parameter $\psi$ determines
a compromise between numerical stability and convergence speed; larger
values of $\psi$ improve the convergence
speed.\footnote{{In Algorithm~\ref{alg: IPI} the range for
    $\psi$ has been empirically determined during our numerical studies.}}


\begin{algorithm}
\KwIn{Matrix $H$; max iterations
      $\subscr{\max}{iter}$; $\psi \in (0.5,1)$.}

    \KwOut{{$(\sigma,\,  z)$} satisfying~\eqref{eq: generalized eig}, or
      \text{fail}.}
    \Repeat{\text{convergence} or $i > \subscr{\max}{iter}$}{
      $ z \gets (H-\mu D)^{-1} D z $\;
      $ {\phi} \gets \| z \| $\;
      $z \gets z/{\phi}$\;
      $\mu = \psi \cdot \min \setdef{{\phi} \in \text{spec}(H,D)}{{\phi} >
  0}$\;
      update $D$ according to~\eqref{eq: def S,T,Q}\;
      $i \gets i+1$    
}
\Return{$( {\phi} + \mu, z)$ or \text{fail} if $i = \subscr{\max}{iter}$}\;
\caption{Heuristic solution to~\eqref{eq: generalized eig}}
\label{alg: IPI}
\end{algorithm}

{When convergent, Algorithm~\ref{alg: IPI} finds a
  solution to~\eqref{eq: generalized eig} and, consequently, the
  algorithm could stop at a local minimum and return a (sub)optimal
  network perturbation preventing observability of a desired
  eigenvalue.} All information about the network matrix, the sensor
nodes, the constraint graph, and the unobservable eigenvalue is
encoded in the matrix $H$ as in \eqref{eq: bar system def},~\eqref{eq:
  optimality} and~\eqref{eq: generalized eig}.  Although convergence
of Algorithm~\ref{alg: IPI} is not guaranteed, numerical studies show
that it performs well in practice; see Sections \ref{sec: line 3}
and~\ref{sec: line star}.

\subsection{Optimal perturbations and algorithm validation}\label{sec:
  line 3}
In this section we validate Algorithm~\ref{alg: IPI} on a small
network. We start with the following result.
\begin{theorem}{\bf \emph{(Optimal perturbations of 3-dimensional line
      networks with fixed $\lambda \in \mathbb{C}$)}}\label{thm: 3 line}
  Consider a network with graph $\mc G = (\mc V , \mc E)$, where
  $|\mc V| = 3$, weighted adjacency matrix
  \begin{align*}
    A=
    \begin{bmatrix}
      a_{11} & a_{12} & 0 \\
      a_{21} & a_{22} & a_{23} \\
      0 & a_{32} & a_{33}
    \end{bmatrix}
                   ,
  \end{align*}
  and sensor node $\mc O = \{1\}$. Let $B = [b_{ij}]= A+ \Delta^*$,
  where $\Delta^*$ solves the minimization problem~\eqref{eq:
    minimization problem F} with {constraint graph}
  $\mc H = \mc G$ and {unobservable eigenvalue}
  $\lambda = \lambda_\Re+\im \lambda_\Im \in \mathbb{C}$,
  $\lambda_\Im \neq 0$.
  Then:
  \begin{align*}
&    b_{11} = a_{11}, \quad b_{21} = a_{21}, \quad b_{12} = 0, \quad 
\end{align*}
and $b_{22}$, $ b_{23}$, $b_{32}$, and $ b_{33}$ satisfy:
  \begin{align}\label{eq: line3x3analyticalequations}
  \begin{split}
    (b_{22} - a_{22}) - (b_{33} - a_{33}) + \frac{b_{33} - b_{22}}{b_{32}} (b_{23} - a_{23})  &=0 , \\
    (b_{32} - a_{32}) -\frac{b_{23}}{b_{32}} (b_{23} - a_{23}) &=0 ,
    \\
    {
    b_{22}+b_{33} - 2 \lambda_{\Re}}&=0 , \\
  {
    b_{22}b_{33}-b_{23}b_{32} - \lambda_{\Re}^2 - \lambda_{\Im}^2} &=0.
    \end{split}
  \end{align}
\end{theorem}

\begin{IEEEproof}
  Let $B x = \lambda x$ and notice that, because $\lambda$ is
  unobservable, $C_{\mc O} x = [1 \; 0 \; 0] x = 0$. Then,
  $x = [x_1 \; x_{2}\; x_3]^\transpose$, $x_1 = 0$, $b_{11} = a_{11}$,
  and $b_{21} = a_{21}$. By contradiction, let $x_2 = 0$.  Notice that
  $Bx = \lambda x$ implies $b_{33} = \lambda$, which contradicts the
  assumption that $\lambda_\Im \neq 0$ and $b_{33} \in \real$. Thus,
  $x_2 \neq 0$. Because $x_2 \neq 0$, the relation $Bx = \lambda x$
  and $x_1 = 0$ imply $b_{12} = 0$. Additionally, $\lambda$ is an
  eigenvalue of
\begin{align*}
  B_2=
  \begin{bmatrix}
    b_{22} & b_{23}\\
    b_{32} & b_{33}
  \end{bmatrix} .
\end{align*}
The characteristic polynomial of $B_2$ is
\begin{align*}
  P_{B_{2}}(s) = s^2 - (b_{22} + b_{33}) s + b_{22} b_{33} - b_{23}b_{32}.
\end{align*}
For $\lambda \in \text{spec} (B_2)$, {we must have
  \mbox{$P_{B_{2}}(s) = (s - \lambda)(s - \bar \lambda)$}, where
  $\bar \lambda$ is the complex conjugate of $\lambda$. Thus,
\begin{align*}
  P_{B_{2}}(s) &= (s - \lambda_\Re - \im \lambda_\Im) (s - \lambda_\Re
                 + \im \lambda_\Im) 
                 = s^2 - 2 \lambda_\Re s + \lambda_\Re^2 + \lambda_\Im^2,
\end{align*}
which leads to
\begin{align}\label{eq: line3x3parameters}
  \begin{split}
    b_{22} + b_{33} - 2 \lambda_\Re = 0 , \text{ and } b_{22} b_{33} -
    b_{23}b_{32} - \lambda_\Re^2 - \lambda_\Im^2 = 0 .
  \end{split}
\end{align}}
The Lagrange function of the minimization problem with cost function
$\| \Delta^* \|_\text{F}^2 = \sum_{i=2}^3 \sum_{j=2}^3 (b_{ij} -
a_{ij})^2$ and constraints~\eqref{eq: line3x3parameters} is
\begin{align*}
\begin{split}
\mc L (b_{22}&, b_{23}, b_{32}, b_{33}, p_1, p_2) = d_{22}^2+d_{23}^2+d_{32}^2+d_{33}^2 \\
&+p_1(2 \lambda_{\Re}+b_{22}+b_{33})
+p_2( b_{22}b_{33}-b_{23}b_{32}-(\lambda_{\Re}^2+\lambda_{\Im}^2)),
\end{split}
\end{align*}
where $p_1,p_2\in \real$ are Lagrange multipliers, and
$d_{ij} = b_{ij} - a_{ij}$. By equating the partial derivatives of
$\mc L$ to zero we obtain
 \begin{align}
 \frac{\partial \mc L}{\partial b_{22}} = 0 &\Rightarrow
     2 d_{22} + p_1 +p_2 b_{33} = 0 ,\label{eq: line(1)} \\
    \frac{\partial \mc L}{\partial b_{33}} = 0 &\Rightarrow
     2 d_{33} + p_1 +p_2 b_{22} = 0 ,\label{eq: line(2)}\\
      \frac{\partial \mc L}{\partial b_{23}} = 0 &\Rightarrow
     2 d_{23} - p_2 b_{32} = 0 , \label{eq: line(3)}\\
           \frac{\partial \mc L}{\partial b_{32}} = 0 &\Rightarrow
     2 d_{32} - p_2 b_{23} = 0 , \label{eq: line(4)}
  \end{align}
  together with~\eqref{eq: line3x3parameters}. The statement follows
  by substituting the Lagrange multipliers $p_1$ and $p_2$ into
  ~\eqref{eq: line(1)} and~\eqref{eq: line(4)}.
\end{IEEEproof}

\begin{figure}[tb]
    \centering
    \includegraphics[width=\columnwidth]{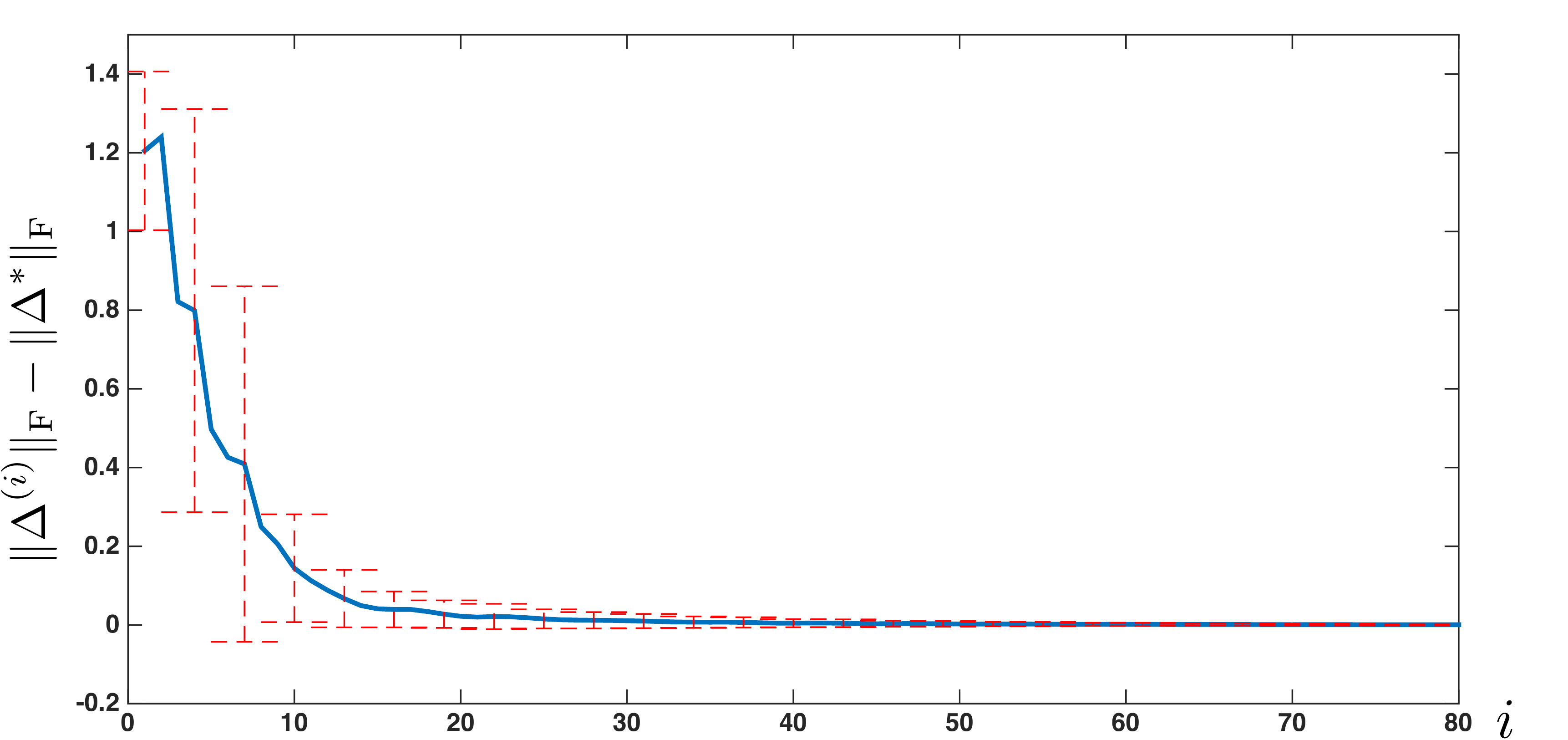}
    \caption{This figure validates the effectiveness of Algorithm
      ~\ref{alg: IPI} to compute optimal perturbations for the line
      network in Section~\ref{sec: line 3}. The plot shows the mean
      and standard deviation over 100 networks of the difference
      between $\Delta^*$, obtained via the optimality conditions
     ~\eqref{eq: line3x3analyticalequations}, and $\Delta^{(i)}$,
      computed at the $i$-th iteration of Algorithm~\ref{alg:
        IPI}. The unobservable eigenvalue is $\lambda = \im$ and the values $a_{ij}$ are chosen independently and
      uniformly distributed in $[0,1]$.}
  \label{fig: 3x3linetrend}
\end{figure}

To validate Algorithm~\ref{alg: IPI}, in Fig.~\ref{fig: 3x3linetrend}
we compute optimal perturbations for 3-dimensional line networks based
on Theorem~\ref{thm: 3 line}, and compare them with the perturbation
obtained at with Algorithm~\ref{alg: IPI}.

\section{{Observability Radius of Random Networks: the Case of Line and
  Star Networks}}\label{sec: line star} In this section we study the
observability radius of networks with fixed structure and random
weights, when the desired unobservable eigenvalue is an optimization
parameter as in \eqref{eq: minimization problem}. First, we give a
general upper bound on the size of an optimal perturbation. Next, we
explicitly compute optimal perturbations for line and star networks,
showing that their robustness is essentially different.

We start with some necessary definitions.  Given a directed graph
$\mc G = (\mc V, \mc E)$, a cut is a subset of edges
$\bar {\mc E} \subseteq \mc E$.  Given two disjoint sets of vertices
$\mc S_1, \mc S_2 \subset \mc V$, we say that a cut $\bar{\mc E}$
disconnects $\mc S_2$ from $\mc S_1$ if there exists no path from any
vertex in $\mc S_2$ to any vertex in $\mc S_1$ in the subgraph
$(\mc V, \mc E\setminus \bar {\mc E})$. Two cuts $\mc E_1$ and
$\mc E_2$ are disjoint if they have no edge in common, that is, if
$\mc E_1 \cap \mc E_2 = \emptyset$.  Finally, the Gamma function is
defined as 
$\Gamma (z) = \int_0^\infty x^{z-1} e^{-x} \; \dd x $.
With this notation
in place, we are in the position to prove a general upper bound on the
(expected) norm of the smallest perturbation that prevents
observability. The proof is based on the following intuition: a
perturbation that disconnects the graph prevents observability.


\begin{theorem}{\bf \emph{(Bound on expected network observability radius)}}\label{thm:
    bound delta} Consider a network with graph
  $\mc G= (\mc V, \mc E)$, weighted adjacency matrix $A = [a_{ij}]$,
  and sensor nodes $\mc O \subseteq \mc V$.  Let the weights $a_{ij}$
  be independent random variables uniformly distributed in the
  interval $[0,1]$. Define the minimal observability-preventing
  perturbation as
    \begin{align}\label{eq:delta-def}
  \delta= \min\limits_{\lambda \in \mathbb{C}, x \in \mathbb{C}^n,  \Delta \in \real^{n\times n}} &  \| \Delta \|_\text{F} ,\\[.7em] 
\nonumber  \text{s.t.} \qquad\qquad  \qquad\qquad             & (A+\Delta) x =
                                                                                \lambda
                                                                                x,\\[.5em]
\nonumber                                                                              &
                                                                                \|
                                                                                x\|_2
                                                                                = 1,\\[.5em]
\nonumber                                                                              &
                                                                                C_{\mc O} x = 0, \\[.5em]
\nonumber                                                                 &
                                                                                \Delta \in \mc A_{\mc G} .
  \end{align}
  Let $\Omega_k(\mc O)$ be a collection of disjoint
  cuts of cardinality $k$, where each cut disconnects a non-empty
  subset of nodes from $\mc O$. Let \mbox{$\omega=|{\Omega_k(\mc O)}|$} be
  the cardinality of $\Omega_k(\mc O)$. Then,
  \begin{align*}
    \E[\delta] 
                    &\le 
                    \frac{\Gamma (1/k)\, \Gamma(\omega+1)}{\sqrt{k} \,\Gamma (\omega + 1 + 1/k)}.
  \end{align*}   
\end{theorem}
\begin{IEEEproof}
  Let $\bar{\mc E} \in \Omega_k(\mc O)$.  Notice
  that, 
  after removing the edges $\bar{\mc E}$, the nodes are partitioned as
  $\mc V = \mc V_1 \cup \mc V_2$, where
  $\mc V_1 \cap \mc V_2 = \emptyset$, $\mc O \subseteq \mc V_1$, and
  $\mc V_2$ is disconnected from $\mc V_1$. Reorder the network nodes
  so that $\mc V_1 = \until{| \mc V_1 |}$ and
  $\mc V_2 = \{| \mc V_1 |+1 , \dots, |\mc V|\}$. Accordingly, the
  modified network matrix is reducible and reads as
  \begin{align*}
    \bar A =
    \begin{bmatrix}
      A_{11} & 0\\
      A_{21} & A_{22}
    \end{bmatrix} .
  \end{align*}
  Let $x_2$ be an eigenvector of $A_{22}$ with corresponding
  eigenvalue $\lambda$. Notice that $\lambda$ is an eigenvalue of
  $\bar A$ with eigenvector $x = [0 \; x_2^\transpose]^\transpose$.
  Since $\mc O \subseteq \mc V_1$, $C_{\mc O} x = 0$, so that the
  eigenvalue $\lambda$ is unobservable.

  From the above discussion we conclude that, for each
  $\bar{\mc E}\in \Omega_k(\mc O)$, there exists a perturbation
  $\Delta = [\delta_{ij}]$ that is compatible with $\mc G$ and ensures
  that one eigenvalue is unobservable. Moreover, the perturbation
  $\Delta$ is defined as $\delta_{ij} = - a_{ij}$ if
  $(i,j) \in \bar{\mc E}$, and $\delta_{ij} = 0$ otherwise. We thus
  have
  \begin{align*}
    \E[ \delta] \le \E \left[ \min_{\bar{\mc E }\in \Omega_k (\mc O)} \sqrt{\sum_{ (i,j) \in \bar{\mc E}} a_{ij}^2 }
    \right].
  \end{align*}
  Because any two elements of $ \Omega_{k} (\mc O)$ have empty
  intersection and all edge weights are independent, we have
  \begin{align*}
    &\text{Pr} \left( \min_{\bar{\mc E} \in \Omega_k(\mc O) } \sqrt{
      \sum_{ (i,j) \in \bar{\mc E}} a_{ij}^2 } \ge x \right) 
      = \text{Pr} \left( \sqrt{\sum_{ (i,j) \in \bar{\mc E}} a_{ij}^2 } \ge x \right)^{\omega}\\
    &= \text{Pr} \left( \sum_{ (i,j) \in \bar{\mc E}} a_{ij}^2  \ge x^2
      \right)^{\omega} 
  =\left(1 - \text{Pr} \left( \sum_{ (i,j) \in \bar{\mc E}} a_{ij}^2  \le x^2
      \right) \right)^{\omega} \!\!.
  \end{align*}
  In order to obtain a more explicit expression for this probability,
  we resort to using a lower bound. Let $a$ denote the vector of
  $a_{ij}$ with $(i,j)\in \bar{\mc E}$. The condition
  $\sum_{ (i,j) \in \bar{\mc E}} a_{ij}^2 \le x^2$ implies that $a$
  belongs to the $k$-dimensional sphere of radius $x$ (centered at the
  origin).  In fact, since $a$ is sampled in $[0,1]^k$, it belongs to
  the intersection between the sphere and the first orthant. By
  computing the volume of the $k$-dimensional cube inscribed in the
  sphere, we obtain
  \begin{align*}
    \text{Pr} \left( \sum_{ (i,j) \in \bar{\mc E}} a_{ij}^2  \le x^2
    \right) \ge 
    \begin{cases} \frac{\left( 2x/\sqrt{k}\right)^k}{2^k} = \left( \frac{x}{\sqrt{k}} \right)^{k}, & \text{ $x\le \sqrt{k}$},\\
    1, & \text{otherwise}.
    \end{cases}
  \end{align*}
  Since $\delta$ takes on nonnegative values only, its expectation can
  be computed by integrating the survival function
    \begin{align*}
      \E[\delta]=\int_0^\infty   \text{Pr} \left(  \delta \ge t \right) \dd t,
    \end{align*}
    which leads us to obtain, {  by suitable changes of variables,
     \begin{align*}
       \E[\delta] &\le \int_0^{\sqrt{k}} \!\left( 1 - \left(
                    \frac{x}{\sqrt{k}} \right)^{k} \right)^{\omega }
                    \!\!\dd x
                    = \sqrt{k}\int_0^1 \!\left( 1 - t^{k} \right)^{\omega }
                    \!\dd t\\
                  &= \frac1{\sqrt{k}}\int_0^1 \!\left( 1 - z \right)^{\omega} z^{\frac1k-1}
                    \dd z = 
                    \frac{1}{\sqrt{k}} \frac{\Gamma (1/k) \Gamma(\omega+1)}{\Gamma (\omega + 1/k + 1)},
                   %
  \end{align*}
  where the last equality follows from the definition of the Beta
  function, $B(x,y) = \int_0^1 t^{x-1} (1-t)^{y-1} \dd t $ for
  \mbox{$\text{Real}(x)>0$}, \mbox{$\text{Real}(y) >0$}, and its
  relation with the Gamma function,
  $B(x,y) = \frac{\Gamma(x) \, \Gamma(y)}{\Gamma(x+y)} $.  }
\end{IEEEproof}
\smallskip
%
We now use Theorem~\ref{thm: bound delta} to investigate the
asymptotic behavior of the expected observability radius on sequences
of networks of increasing cardinality $n$.  In order to emphasize the
dependence on $n$, we shall write $\E [\delta(n)]$ from now on. As a
first step, we can apply Wendel's inequalities~\cite{JGW:48} to find
\begin{align*}
  \frac{1}{(\omega +1)^{1/k}} \le \frac{ \Gamma(\omega+1)}{\Gamma
                    (\omega + 1 + 1/k)}\le \frac{(\omega +1 + 1/k)^{1-1/k}}{
  (\omega +1)} .
\end{align*}
If in a sequence of networks $\omega$ grows to infinity and $k$
remains constant, then the ratio between the lower and the upper bounds goes to one, yielding the asymptotic equivalence
\begin{align*}
  \E [\delta(n)] \le \frac{\Gamma (1/k)\, \Gamma(\omega+1)}{\sqrt{k} \,\Gamma
                    (\omega + 1 + 1/k)} \sim \frac{\Gamma (1/k)}{
  \sqrt{k}} \frac1{(\omega +1)^{1/k}}.
\end{align*}
This relation implies that a network becomes less robust to
perturbations as the size of the network increases, with a rate
determined by $k$.
In the rest of this section we study two network topologies with
different robustness properties. In particular, we show that line
networks achieve the bound in Theorem~\ref{thm: bound delta}, proving
its tightness, whereas star networks have on average a smaller
observability~radius.

\begin{figure}[t]
  \centering \subfigure[Line network]{
    \includegraphics[width=.42\columnwidth]{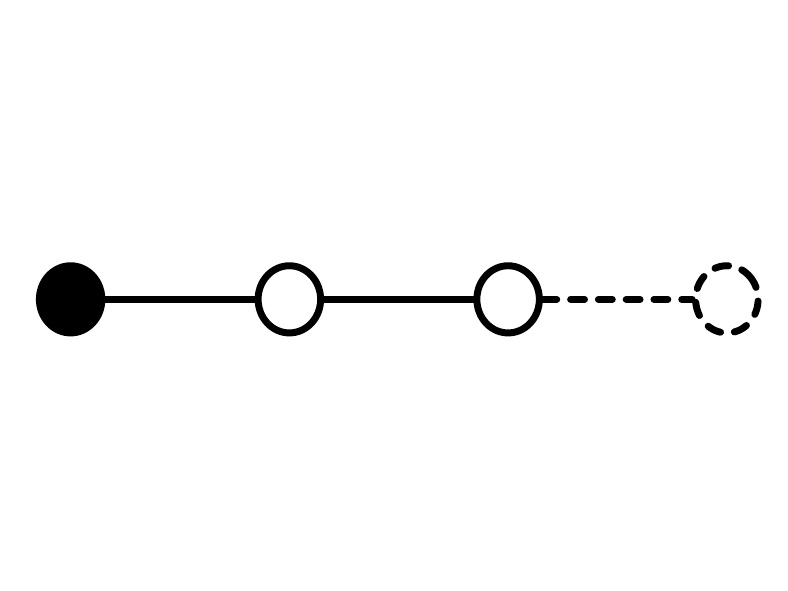}
    \label{fig: Line&Star-Line}
  } \;\;\subfigure[Star network]{
    \includegraphics[width=.42\columnwidth]{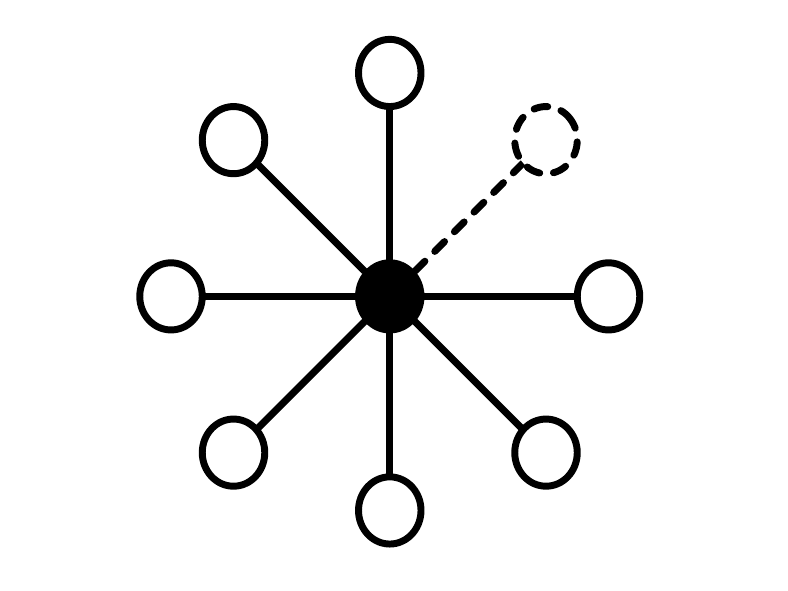}
    \label{fig: Line&Star-Star}
  }
  \caption{Line and star networks with self-loops. Sensor nodes are marked in black. Note that, self-loops are not shown.}
  \label{fig: Line&Star}
\end{figure}

\smallskip
\noindent
\emph{(Line network)} Let $\mc G$ be a line network with $n$ nodes and
one sensor node as in Fig.~\ref{fig: Line&Star}. The adjacency and
output matrices read as
\begin{align}\label{eq: A and B line}
  \begin{split}
      A &= 
    \begin{bmatrix}
      a_{11} & a_{12} & 0 & \cdots & 0\\
      a_{21} & a_{22} & a_{23} & \cdots & 0\\
      \vdots & \ddots & \ddots & \ddots & \vdots\\
      0 & \cdots & a_{n-1,n-2} & a_{n-1,n-1} & a_{n-1,n}\\
      0 & \cdots & 0 & a_{n,n-1} & a_{nn}
    \end{bmatrix}
                                   ,\\
                                   C_{\mc O} &=
                                   \begin{bmatrix}
                                     1 & 0 & 0 & \cdots & 0
                                   \end{bmatrix}.
  \end{split}
\end{align}
We obtain the following result.
\begin{theorem}{\bf \emph{(Structured perturbation of line
      networks)}}\label{thm: resilience line}
  Consider a line network with matrices as in~\eqref{eq: A and B
    line}, where the weights $a_{ij}$ are independent random variables
  uniformly distributed in the interval $[0,1]$. Let $\delta(n)$ be
  the minimal cost defined as in~\eqref{eq:delta-def}.
  Then,
  \begin{align*}
    &\delta (n) = \min \{ a_{12}, \dots, a_{n-1,n}\}, \text{ and }
    \E [\delta(n)] = \frac1n .
  \end{align*}
\end{theorem}
\begin{IEEEproof}
  It is known that line networks, when observed from one of their
  extremes, are strongly structurally observable, that is, they are
  observable for every nonzero choice of the edge
  weights~\cite{HM-TY:79}. Consequently, for the perturbed system to
  feature an unobservable eigenvalue, the perturbation $\Delta$ must
  be such that $\delta_{i,i+1} = -a_{i,i+1}$ for some
  $i \in \{ 2,\dots, n-1\}$.
  Thus, a minimum norm perturbation is obtained by selecting the
  smallest entry $a_{i,i+1}$.
  Since the $a_{i,i+1}$ are independent and identically distributed,
  $\delta(n) = \min a_{i,i+1}$ is a random variable with survival
  function $\text{Pr} (\delta(n) \ge x) = (1-x)^{n-1}$ for
  $0 \le x \le 1$, and $\text{Pr} (\delta(n) \ge x) = 0$
  otherwise. Thus,
  \begin{align*}
    \E[\delta(n)] &= \int_0^1 \text{Pr} (\delta(n) \ge x) \dd x = \frac{1}{n}.
      \end{align*}
%
\end{IEEEproof}

Theorem~\ref{thm: resilience line} characterizes the resilience of
line networks to structured perturbations. 
We remark that, because line networks are strongly structurally
observable, structured perturbations preventing observability
necessarily disconnect the network by zeroing some network weights.
Consistently with this remark, line networks achieve the upper bound
in Theorem~\ref{thm: bound delta}, being therefore maximally robust to
structured perturbations. In fact, for $\mc O = \{1\}$ and a cut size
$k = 1$ we have $\Omega_1 (\mc O) = \{a_{12},\dots, a_{n-1,n}\}$ and
$\omega = n-1$.
Thus, 
\begin{align*}
  \E [\delta (n)] \le \frac{\Gamma (1) \Gamma(n)}{\sqrt{1} \Gamma
  (n+1)} = \frac{(n-1)!}{n!} = \frac{1}{n} ,
\end{align*}
which equals the behavior identified in Theorem~\ref{thm: resilience
  line}. {Further, Theorem~\ref{thm: resilience line} also
  identifies an unobservable eigenvalue yielding a perturbation with
  minimum norm. In fact, if
  \mbox{$a_{i^*-1 , i^*} = \min \{a_{12}, \dots, a_{n-1, n}\}$}, then all
  eigenvalues of the submatrix of $A$ with rows/columns in the set
  $\{i^*,\dots, n\}$ are unobservable, and thus minimizers in
 ~\eqref{eq:delta-def}.} 

Both Theorems~\ref{thm: bound delta} and~\ref{thm: resilience line}
are based on constructing perturbations by disconnecting the
graph. This strategy, however, suffers from performance limitations
and may not be optimal in general.
The next example shows that different kinds of perturbations, when
applicable, may yield a lower cost.


\smallskip
\noindent
\emph{(Star network)} Let $\mc G$ be a star network with $n$ nodes and
one sensor node as in Fig.~\ref{fig: Line&Star}. The adjacency and
output matrices read as
\begin{align}\label{eq: A and B star}
  \begin{split}
      A &= 
    \begin{bmatrix}
      a_{11} & a_{12} & a_{13} & \cdots & a_{1n}\\ 
      a_{21} & a_{22} & 0 & \cdots & 0\\
      a_{31} & 0 & \ddots & \ddots & \vdots\\
      \vdots & \vdots & 0 & a_{n-1,n-1} & 0\\
      a_{n1} & 0 & 0 & 0 & a_{nn}
    \end{bmatrix}
                                   ,\\
                                   C_{\mc O} &=
                                   \begin{bmatrix}
                                     1 & 0 & 0 & \cdots & 0
                                   \end{bmatrix}.
  \end{split}
\end{align}
Differently from the case of line networks, star networks are not
strongly structurally observable, so that different perturbations may
result in unobservability of some modes.

\begin{theorem}{\bf \emph{(Structured perturbation of star
      networks)}}\label{thm: star}
  Consider a star network with matrices as in~\eqref{eq: A and B
    star}, where the weights $a_{ij}$ are independent random variables
  uniformly distributed in the interval $[0,1]$. Let $\delta(n)$ be the
  minimal cost defined as in~\eqref{eq:delta-def}.
  Let
  \begin{align*}
    \gamma = \min_{i,j \in \{ 2,\dots,n\},i\neq j} \frac{| a_{ii} -
    a_{jj} |}{ \sqrt{2}} .
  \end{align*}
  Then,
  \begin{align*}
    \delta (n) = \min \{ a_{12}, a_{13}, \dots, a_{1n} , \gamma\}, \text{ and }\\[.3em]
\frac{1}{\sqrt{2} \, n (n-1)}\le     \E [\delta(n)] \le  \frac{1}{\sqrt{2} \, n (n-2)}.
  \end{align*}
\end{theorem}
\smallskip
\begin{IEEEproof}
  Partition the network matrix $A$ in~\eqref{eq: A and B star} as
  \begin{align*}
    A =
    \begin{bmatrix}
      a_{11} & A_{12}\\
      A_{21} & A_{22}
    \end{bmatrix}
                        ,
  \end{align*}
  where $A_{12} \in \real^{1,n-1}$, $A_{21} \in \real^{n-1,1}$,
  $A_{22} \in \real^{n-1,n-1}$. Accordingly, let
  $x = [x_1 \; x_2^\transpose]^\transpose$. The condition
  $C_{\mc O} x = 0$ implies $x_1 = 0$. Consequently, for the condition
  $(A+\Delta )x = \lambda x$ to be satisfied, we must have
  $(A_{12} + \Delta_{12}) x_2 = 0$ and
  $(A_{22} + \Delta_{22}) x_2 = \lambda x_2$. Notice that, because
  $A_{22}$ is diagonal and $\Delta \in \mc A_{\mc G}$, the condition
  $(A_{22} + \Delta_{22}) x_2 = \lambda x_2$ implies that
  $\lambda = a_{ii} + \delta_{ii}$ for all indices $i$ such that
  $i \in \text{Supp}(x_2)$, where $\text{Supp}(x_2)$ denotes the set
  of nonzero entries of $x_2$. Because $\| x\| = 1$,
  $|\text{Supp}(x_2)| > 0$. We have two cases:

  \smallskip
  \emph{Case $|\text{Supp}(x_2)| = 1$:} Let
  $\text{Supp}(x) = \{ i \}$, with $i \in \{2,\dots,n\}$. Then, the
  condition $(A_{12} + \Delta_{12}) x_2 = 0$ implies
  $\delta_{1,i} = - a_{1,i}$, and the condition
  $(A_{22} + \Delta_{22}) x_2 = \lambda x_2$ is satisfied with
  $\Delta_{22} = 0$, $\lambda = a_{ii}$, and $x = e_i$, where $e_i$ is
  the $i$-th canonical vector of dimension $n$. Thus, if
  $|\text{Supp}(x_2)| = 1$, then
  $\delta(n) = \min_{i \in \{2, \dots, n \}} a_{1,i}$.

  \smallskip \emph{Case $|\text{Supp}(x_2)| > 1$:} Let
  $S = \text{Supp}(x_2)$. Then, $\delta_{ii} = \lambda - a_{ii}$.
  Notice that the condition $(A_{22} + \Delta_{22}) x_2 = \lambda x_2$
  is satisfied for every $x_2$ with support $S$ and, particularly, for
  $x_2 \in \Ker(A_{12})$. Thus, we let $\Delta_{12} = 0$. Notice that
  \begin{align*}
    \delta(n) = \min_{\lambda, S} \sqrt{\sum_{ i \in S} (\lambda - a_{ii}
    )^2} ,
  \end{align*}
  and that $\delta(n)$ is obtained when $S = \{i , j\}$, for some
  $i,j \in \{2, \dots, n \}$, and $\lambda = (a_{ii} + a_{jj})/2$.
  Specifically, for the indexes $\{i, j\}$, we have
  $\| \Delta \|_\text{F} = | a_{ii} - a_{jj}| / \sqrt{2}$. Thus, if
  $|\text{Supp}(x_2)| > 1$, then $\delta(n) = \gamma$, which concludes
  the proof of the first statement.

In order to estimate $\E[\delta(n)]$, notice that $\delta(n) = \min \{\alpha , \gamma\}$,
  where $\alpha = \min \{ a_{12}, a_{13},\dots, a_{1n} \}$, and that
  $\alpha$ and $\gamma$ are independent random variables. Then, from~\cite[Chapter
  6.4]{HAD-HNN:03} we have
  \begin{align*}
    \text{Pr}(\delta(n) \ge x) &= \text{Pr}(\alpha \ge x)
                                 \text{Pr}(\gamma \ge x) \\ 
                               &= (1-x)^{n-1} (1-
                                 (n-2)\sqrt{2}x)^{n-1} ,
  \end{align*}
  for $x \le (\sqrt{2}(n-2))^{-1}$, and
  $\text{Pr}(\delta(n) \ge x) = 0$ otherwise. Thus,
  \begin{align*}
    \E[\delta(n)] = \int_0^{\frac{1}{\sqrt{2}(n-2)}} (1-x)^{n-1} (1-
    (n-2)\sqrt{2}x)^{n-1} \dd x. 
  \end{align*}
  Next, for the upper bound observe that 
\begin{align*}
&\int_0^{\frac{1}{\sqrt{2}(n-2)}} (1-x)^{n-1} (1-
    (n-2)\sqrt{2}x)^{n-1} \dd x\\&\le
    \int_0^{\frac{1}{\sqrt{2}(n-2)}} (1-
    (n-2)\sqrt{2}x)^{n-1} \dd x=\frac1{\sqrt{2}n(n-2)},
\end{align*}
and for the lower bound observe that 
\begin{align*}
&\int_0^{\frac{1}{\sqrt{2}(n-2)}} (1-x)^{n-1} (1-
    (n-2)\sqrt{2}x)^{n-1} \dd x
\\    &=
    \int_0^{\frac{1}{\sqrt{2}(n-2)}} (1-((n-2)\sqrt{2}+ 1) x+ ((n-2)\sqrt{2}) x^2)^{n-1} \dd x
    \\&\ge
    \int_0^{\frac{1}{\sqrt{2}(n-1)}} (1-
    (n-1)\sqrt{2}x)^{n-1} \dd x=\frac1{\sqrt{2}n(n-1)}.
\end{align*}
\end{IEEEproof}


\begin{figure}
    \centering
    \includegraphics[width=\columnwidth]{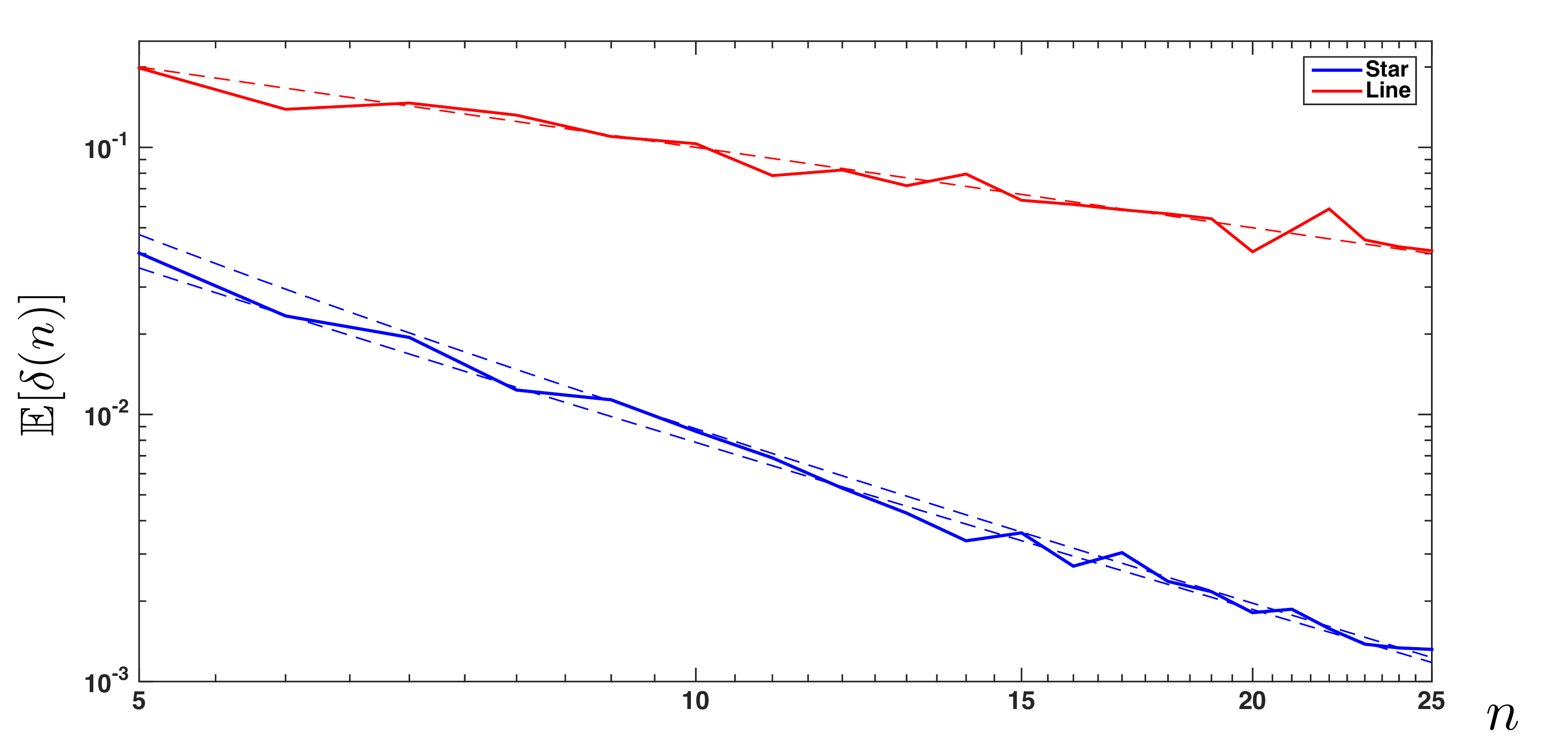}
    \caption{
      Expected values $\E [ \delta(n) ]$ for the two network
      topologies in Fig.~\ref{fig: Line&Star} as functions of the
      network cardinality $n$. Dotted lines represent upper and lower
      bounds in Theorems~\ref{thm: resilience line} and~\ref{thm:
        star}.  Solid lines show the mean over 100 networks of the
      Frobenius norm of the perturbations obtained by
      Algorithm~\ref{alg: IPI}.}
  \label{fig: frobNorm&NetSize}
\end{figure}

{Theorem~\ref{thm: star} quantifies the resilience of star
  networks, and the unobservable eigenvalues requiring minimum norm
  perturbations; see the proof for a characterization of this
  eigenvalues.}

The bounds in Theorem~\ref{thm: star} are asymptotically tight and
imply 
\begin{align*}
  \E [\delta (n)]\sim \frac1{\sqrt{2}\, n^2}, \quad
  {\text{as }n\to\infty}.
\end{align*}
See Fig.~\ref{fig: frobNorm&NetSize} for a numerical validation of
this result.  This rate of decrease implies that star networks are
\emph{structurally} less robust to perturbations than line
networks. Crucially, unobservability in star networks may be caused by
two different phenomena: the deletion of an edge disconnecting a node
from the sensor node (deletion of the smallest among the edges
{$\{a_{12},a_{13},\dots, a_{1n}\}$}), and the creation of
a dynamical symmetry with respect to the sensor node by perturbing two
diagonal elements to make them equal in weight.  It turns out that, on
average, creating symmetries is ``cheaper'' than disconnecting the
network. The role of network symmetries in preventing observability
and controllability has been observed in several independent works;
see for instance \cite{AC-MM:14,FP-SZ:14}. Finally, the comparison of
line and star networks shows that Algorithm~\ref{alg: IPI} is a useful
tool to systematically investigate the robustness of different
topologies.


\section{Conclusion}\label{sec: conclusion}
In this work we {extend the notion of observability radius
  to network systems, thus providing a measure of the ability to
  maintain observability} of the network modes against structured
perturbations of the edge weights. We 
characterize network perturbations preventing observability, and
describe a heuristic algorithm to compute perturbations with smallest
Frobenius norm. Additionally, we study the observability radius of
networks with random weights, derive a fundamental bound relating the
observability radius to certain connectivity properties, and
explicitly characterize the observability radius of line and star
networks. Our results show that different network structures
{exhibit inherently different robustness properties}, and
thus provide guidelines for the design of robust complex
networks. 




\bibliographystyle{IEEEtran}
\bibliography{./BIB}

\end{document}